\documentclass{article}%
\usepackage{amssymb}
\usepackage{amsmath}%
\usepackage{amsfonts}%
\usepackage{graphicx}

\numberwithin{equation}{section}
\begin{document}

\title{The diffusive compressible Euler model in moving reference frames}
\author{M. L. Morris\\independent reseacher\\Albuquerque, New Mexico, USA\\melissamorris443@gmail.com}
\date{October 25, 2024}
\maketitle

\begin{abstract}
We continue to investigate the diffusive compressible Euler (dcE) model for
viscous and heat conducting compressible fluid flow, which has been proposed
by M. Sv\"{a}rd as an alternative to the Navier-Stokes-Fourier (NSF)
equations. The non-convective contribution to the momentum flux tensor in the
dcE model is, with inverted sign, the analog of the viscous stress tensor in
the NSF formulation. Unlike the latter quantity, the former tensor is
non-symmetric, and here we examine some of the consequences of this property.
In particular, we demonstrate that the dcE model's analog viscous stress
tensor, and its resulting analog viscous dissipation term, are not objective
-- that is to say, these quantities, under general time-dependent Euclidean
transformations, feature moving reference frame dependence.

\end{abstract}

\section{\label{intro}Introduction}

It is widely recognized -- see \textsc{Segel} \cite[Sec. 3.1]{segel}, for
example -- that the symmetric viscous stress tensor for Newtonian fluids in
the Navier-Stokes-Fourier (NSF) model leads to the objectivity of both its
constitutive law and the viscous dissipation term that arises in the internal
energy and entropy balance laws. That is to say, the viscous stress and
viscous dissipation both transform into a moving reference frame in the
expected way and without any additional contributions from the frame itself.
Because its local mass conservation is governed by the continuity equation,
the NSF model may be cast in a Lagrangian description whereby material volumes
-- or, in infinitesimal limit, material points -- which conserve mass, can be
tracked invertibly throughout the fluid motion, even as the volume may deform
due to the fluid's compressibility. The NSF model is a well-known example of a
formulation that obeys the principle of material frame indifference, which
guarantees that in the Lagrangian setting, rigid body motions superimposed on
its constitutive relations leave them unaltered. However, we will see that the
Lagrangian description is not a requirement when studying matters of
objectivity and reference frame invariance.

Proposed as an alternative to the NSF equations, the diffusive compressible
Euler (dcE) model of \textsc{Sv\"{a}rd} (\cite{sv1}, \cite{sv24}) is based
primarily on the idea that the constitutive relations for its non-convective
fluxes derive from the transport of mass, linear momentum, and total energy by
fluid particles as they diffuse. In the case of the dcE model, this view leads
to a mass diffusion flux included in the mass conservation law and to a
non-symmetric momentum diffusion flux appearing in the momentum balance law.
In \textsc{Sv\"{a}rd} \cite[Sec. 1]{sre} the author suggests that his shift in
paradigm precludes the use of traditional NSF concepts and\ terminology, such
as viscosity, when describing fluids in the dcE model. Nonetheless, we feel
that it is necessary for comparison to construct some of the dcE model's
analog quantities to their NSF counterparts and study them side by side. To
this end, in Section \ref{sem} we identify the dcE model's analog to the
viscous stress tensor as the momentum diffusion flux (with inverted sign), and
we derive the internal energy balance law in order to identify the dcE model's
analog to the viscous dissipation heating. In Appendix \ref{apps}, we
additionally use the internal energy equation and the standard framework of
irreversible thermodynamics to derive an entropy balance law for the dcE
model, and we use the resulting expression for its volumetric entropy
production rate to find second law of thermodynamics restrictions.

The topic of study in Section \ref{obj} is related to the aforementioned
principle of material frame indifference -- although, to avoid confusion, we
do not use this term when applied to the dcE model.\ This is because, since
the model in question features mass diffusion in its mass balance law, it
lacks an invertible material point mapping like the one seen in the NSF
equations of fluid flow. This, in turn, leads to the dcE model having no
Lagrangian description in the traditional sense and so material coordinates
may no longer be used to study the issue.\ If one wishes to address matters of
reference frame independence in the dcE model, it must be done using Eulerian
coordinates, a task which fortunately presents no major obstacles.\ In fact,
we will see that some of the surrounding issues are easier to interpret as
mathematical principles anchored to an Eulerian description. Throughout
Section \ref{obj}, we examine certain quantities of interest under a general
time-dependent Euclidean transformation -- that is, a mathematical
transformation that preserves relative distances and angles and, thereby, is
associated with an arbitrary moving reference frame. The quantity is called
objective if, after being transformed into the moving coordinates, there arise
no additional effects from the rotation or translation of the frame, itself.
Galilean invariance is the special case of objectivity in which there is no
rotation and the translation is at a constant velocity. The constitutive laws
for both the NSF and dcE models are investigated in Section \ref{vst}, where
it is shown that all are objective except for the dcE model's analog viscous
stress tensor, which is further shown to not be Galilean invariant. Next, in
Sections \ref{vd} and \ref{vrep}\ we examine the viscous dissipation, which
arises as a production term in the internal energy and entropy balance laws,
to demonstrate this quantity is objective in the NSF model, but its analog in
the dcE model is not.

We wish to emphasize that objectivity is a mathematical property, whereas the
principle of material frame indifference is a physical expectation that
constitutive relations should be objective, and as pointed out in
\textsc{Sv\"{a}rd} \cite[Sec. 5]{sre}, some researchers contest the general
application of this principle. The current discussion, however, deals not in
generalities, but with the specific fluid dynamics models, NSF and dcE, and
here we carry out Euclidean transformations on physical quantities within
these models to determine -- not only whether or not they are objective -- but
the actual frame-dependent terms, if non-objective. By examining these terms,
one is presented with the opportunity to consider whether or not the
particular type of frame dependence exhibited by a quantity may be physically
justified. However, whether or not one is in favor of the requirement that the
constitutive laws, themselves, satisfy objectivity, it is important to
recognize certain quantities that derive from these constitutive laws are,
inarguably, required to be objective. For example, the volumetric entropy
production rate, which is used to compute the total dissipation in a fluid
system, must not vary between a fixed and moving reference frame. Otherwise,
this would have the unphysical implication that reference frame motion affects
the amount of fluid dissipation.

Throughout this discussion, vectors are denoted in bold-face, order $2$
tensors in sans serif font, $\mathsf{I}$\ is used to represent the order\ $2$
identity tensor, and standard tensor operators have been employed as in
\textsc{Bird} \textit{et al}. \cite{bird}, for example. Some additional
definitions and relevant tensor identities are provided in Appendix \ref{ten}
and, so as not to distract from the main ideas of the text, some of the
intermediate computations are carried out there, as well. The results for the
NSF model are well-known and treated in textbooks like \textsc{de Groot} and
\textsc{Mazur} \cite{degroot} and \textsc{Segel} \cite{segel}. We provide the
details of obtaining them here to aid our comparison with the dcE model.

\textsc{Morris} \cite{me}\ is an earlier, more condensed version of the
current discussion in which many of the salient points have already been
addressed. In his rebuttal \textsc{Sv\"{a}rd} \cite[Sec. 5]{sre}, it appears
that the author has either ignored or not fully understood the subject
material presented in \cite{me}. In the event of the latter, we provide
additional details and explanations below and, along the way, we take the
opportunity to dispel some of the false claims made in \cite{sre}.

\section{\label{secNSF}The Navier-Stokes-Fourier model}

When there are no external body forces like gravity included, the NSF model
may be expressed as the following system of mass, momentum, and total energy
conservation laws:%
\begin{align}
\frac{\partial\rho}{\partial t}  & =-\nabla\cdot\left(  \rho\mathbf{v}\right)
,\label{a1}\\
\frac{\partial\left(  \rho\mathbf{v}\right)  }{\partial t}  & =-\nabla
\cdot\left(  p\mathsf{I}-\mathsf{S}+\rho\mathbf{vv}\right)  ,\label{a2}\\
\frac{\partial E}{\partial t}  & =-\nabla\cdot\left(  \mathbf{q}%
-\mathsf{S}\cdot\mathbf{v}+\left(  E+p\right)  \mathbf{v}\right)  ,\label{a3}%
\end{align}
where $\rho$ and $p$\ represent the mass density and thermodynamic pressure,
respectively; $\mathbf{v}$\ denotes the local fluid velocity; and $E$\ is the
total energy density, computed as the sum of the internal energy density, $u$,
and the kinetic energy density:%
\begin{equation}
E=u+\frac{1}{2}\rho\mathbf{v}\cdot\mathbf{v}.\label{a4}%
\end{equation}
Furthermore, $\mathbf{q}$\ and $\mathsf{S}$\ represent the heat flux vector
and the viscous stress tensor, respectively, with constitutive equations given
by Fourier's law:%
\begin{equation}
\mathbf{q}=-\lambda\nabla T,\label{a5}%
\end{equation}
where $T$ represents the absolute temperature and $\lambda$\ is the thermal
conductivity, and Newton's viscosity law:%
\begin{equation}
\mathsf{S}=\zeta\left(  \nabla\cdot\mathbf{v}\right)  \mathsf{I}+2\eta
\nabla\mathbf{v}^{SD},\label{a6}%
\end{equation}
where $\zeta$\ is the bulk viscosity, $\eta$ is the shear viscosity, and the
superscript "$SD$" is used to\ indicate the symmetric deviatoric part of a tensor.

From the NSF conservation equations above, one may derive a few other balance
laws. As computation \ref{com0} in Appendix \ref{com}, it is shown that mass
and momentum conservation equations (\ref{a1}) and (\ref{a2}) lead to the
kinetic energy balance law:%
\begin{equation}
\frac{\partial\left(  \frac{1}{2}\rho\mathbf{v}\cdot\mathbf{v}\right)
}{\partial t}=-\nabla\cdot\left(  -\mathsf{S}\cdot\mathbf{v}+\left(
p+\frac{1}{2}\rho\mathbf{v}\cdot\mathbf{v}\right)  \mathbf{v}\right)
+p\nabla\cdot\mathbf{v}-\Psi,\label{a7}%
\end{equation}
where $\Psi$ symbolizes the viscous dissipation term,%
\begin{equation}
\Psi=\zeta\left(  \nabla\cdot\mathbf{v}\right)  ^{2}+2\eta\nabla
\mathbf{v}^{SD}\cdot\cdot\nabla\mathbf{v}^{SD},\label{f2}%
\end{equation}
and the double-dot operator represents the inner tensor product defined in
(\ref{t9}). By next using total energy conservation equation (\ref{a3}),
together with (\ref{a4}) and (\ref{a7}), one obtains the internal energy
balance law:%
\begin{equation}
\frac{\partial u}{\partial t}=-\nabla\cdot\left(  \mathbf{q}+u\mathbf{v}%
\right)  -p\nabla\cdot\mathbf{v}+\Psi.\label{f1}%
\end{equation}
Note that if the viscosity parameters, $\zeta$ and $\eta$, are non-negative,
then $\Psi$ is guaranteed to be non-negative, as well. As discussed in
\textsc{Bird} \textit{et al}. \cite[p. 82]{bird}, $\Psi$\ is sometimes called
the "viscous dissipation heating" since it accounts for the "degradation of
mechanical energy into thermal energy."

In Appendix \ref{apps}, we use internal energy and mass balance equations,
(\ref{f1}) and (\ref{a1}), to derive the following entropy balance law:%
\begin{equation}
\frac{\partial s}{\partial t}=-\nabla\cdot\left(  -\frac{\lambda}{T}\nabla
T+s\mathbf{v}\right)  +\sigma,\label{ff1}%
\end{equation}
where $s$\ denotes the entropy density and the volumetric entropy production
rate is given by%
\begin{equation}
\sigma=\frac{\lambda}{T^{2}}\nabla T\cdot\nabla T+\frac{\Psi}{T}.\label{ff2}%
\end{equation}
The second law of thermodynamics requires that $\sigma$\ never become
negative, and one observes for the NSF model that this condition is upheld
when each of its transport parameters, $\zeta$, $\eta$, and $\lambda$, is non-negative.

\section{\label{sem}The diffusive compressible Euler model}

The fluid dynamics model proposed in \textsc{Sv\"{a}rd} \cite{sv24} as an
alternative to the NSF equations is the following system of balance laws for
the mass, momentum, and total energy:%
\begin{align}
\frac{\partial\rho}{\partial t}  & =-\nabla\cdot\left(  -\nu\nabla\rho
+\rho\mathbf{v}\right) \label{f3}\\
\frac{\partial\left(  \rho\mathbf{v}\right)  }{\partial t}  & =-\nabla
\cdot\left(  p\mathsf{I}-\nu\nabla\left(  \rho\mathbf{v}\right)
+\rho\mathbf{vv}\right) \label{f4}\\
\frac{\partial E}{\partial t}  & =-\nabla\cdot\left(  -\nu\nabla E-\kappa
_{T}\nabla T+\left(  E+p\right)  \mathbf{v}\right)  ,\label{f5}%
\end{align}
where $\nu$\ and $\kappa_{T}$\ are transport parameters that the author refers
to as the diffusion coefficient and the heat diffusion coefficient, respectively.

In a similar manner as Section \ref{secNSF}, one may use the mass and momentum
conservation equations (\ref{f3}) and (\ref{f4}) to obtain the kinetic energy
balance law for the dcE model:%
\begin{equation}
\frac{\partial\left(  \frac{1}{2}\rho\mathbf{v}\cdot\mathbf{v}\right)
}{\partial t}=-\nabla\cdot\left(  -\nu\nabla\left(  \frac{1}{2}\rho
\mathbf{v}\cdot\mathbf{v}\right)  +\left(  p+\frac{1}{2}\rho\mathbf{v}%
\cdot\mathbf{v}\right)  \mathbf{v}\right)  +p\nabla\cdot\mathbf{v}-\rho
\nu\nabla\mathbf{v}\cdot\cdot\nabla\mathbf{v}.\label{f6.1}%
\end{equation}
As computation \ref{com1} of Appendix \ref{com}, we provide the details of
obtaining Equation (\ref{f6.1}), which one may employ, along with relation
(\ref{a4}) and total energy conservation equation (\ref{f5}), to obtain the
following internal energy balance law in the dcE model:%
\begin{equation}
\frac{\partial u}{\partial t}=-\nabla\cdot\left(  -\nu\nabla u-\kappa
_{T}\nabla T+u\mathbf{v}\right)  -p\nabla\cdot\mathbf{v}+\rho\nu
\nabla\mathbf{v}\cdot\cdot\nabla\mathbf{v}.\label{f7}%
\end{equation}

Upon comparing the NSF and dcE models' momentum conservation equations,
(\ref{a2}) and (\ref{f4}), and internal energy balance laws, (\ref{f1}) and
(\ref{f7}), one observes that the dcE analogs to the NSF viscous stress tensor
(\ref{a6}) and NSF viscous dissipation term (\ref{f2})\ are given by%
\begin{equation}
\mathsf{S}_{dcE}=\nu\nabla\left(  \rho\mathbf{v}\right)  ,\label{f7.5}%
\end{equation}
and%
\begin{equation}
\Psi_{dcE}=\rho\nu\nabla\mathbf{v}\cdot\cdot\nabla\mathbf{v},\label{f8}%
\end{equation}
respectively. Note that if the diffusion coefficient, $\nu$, is non-negative,
then $\Psi_{dcE}$ is also non-negative and thereby represents a dissipative
heating term in the same role as $\Psi$\ from the previous
section.\footnote{{}Please consider from \textsc{Sv\"{a}rd} \cite[Sec. 5]%
{sre}: "...Morris recasts the diffusive terms into something resembling
viscous terms...", and note that in the current discussion and its earlier
condensed version \textsc{Morris} \cite{me} to which the statement refers,
there has been no re-expression of dcE terms to contrive a semblance to
viscosity -- the terms in question arise naturally as analogs from the dcE
model, itself, regardless of whether one refers to them as "viscous", "viscous
analog", "diffusive", or "linear momentum diffusive".}

In Appendix \ref{apps}, we use the internal energy and mass balance equations
(\ref{f7}) and (\ref{f3}) to derive the following entropy balance law for the
dcE model:%
\begin{equation}
\frac{\partial s}{\partial t}=-\nabla\cdot\left(  -\nu\nabla s-\frac
{\kappa_{T}}{T}\nabla T+s\mathbf{v}\right)  +\sigma_{dcE}\label{ff4}%
\end{equation}
with volumetric entropy production rate given by%
\begin{equation}
\sigma_{dcE}=\frac{1}{T^{2}}\left(  \frac{\rho c_{p}\nu}{\gamma}+\kappa
_{T}\right)  \nabla T\cdot\nabla T+\frac{c^{2}\nu}{\rho T\gamma}\nabla
\rho\cdot\nabla\rho+\frac{\Psi_{dcE}}{T}.\label{ff5}%
\end{equation}
In the above, the symbols, $c$, $c_{p}$, and $\gamma$, denote the adiabatic
sound speed, the isobaric specific heat, and the ratio of isobaric to
isochoric specific heat, respectively, all of which are positive quantities.
Therefore, we see that in the dcE model, the second law of thermodynamics
requirement -- i.e., that $\sigma_{dcE}\geq0$ -- is met when the diffusion
coefficient, $\nu$, is non-negative and the thermal diffusion coefficient
satisfies $\kappa_{T}\geq-\rho c_{p}\nu/\gamma$.

\section{\label{obj}Objectivity}

\subsection{\label{et}Time-dependent Euclidean transformations}

In an Eulerian description, let us regard $\mathbf{x}$\ as the position in a
fixed reference frame and $\widehat{\mathbf{x}}$\ as the position in a moving
reference frame. Here, we present this purely as a mathematical construct--the
moving reference frame is arbitrary and need not necessarily be related to a
Lagrangian description. Coordinate mappings between the two frames are
represented by Euclidean transformations, i.e., ones that preserve relative
distances and angles, the most general of which is given by%
\begin{equation}
\widehat{\mathbf{x}}=\mathsf{Q}\left(  t\right)  \cdot\mathbf{x}%
+\mathbf{o}\left(  t\right)  ,\label{f9}%
\end{equation}
where the order $2$\ tensor, $\mathsf{Q}\left(  t\right)  $, is real
orthogonal:%
\begin{equation}
\mathsf{Q}^{T}\cdot\mathsf{Q}=\mathsf{Q}\cdot\mathsf{Q}^{T}=\mathsf{I}%
,\label{f10}%
\end{equation}
the vector, $\mathbf{o}\left(  t\right)  $, is the origin of the moving
reference frame, and both are assumed to be continuously differentiable. The
first and second terms on the right-hand side of (\ref{f9}) are, respectively,
the rotation and translation of the moving reference frame, and the coordinate
mapping is invertible with%
\begin{equation}
\mathbf{x}=\mathsf{Q}^{T}\cdot\left(  \widehat{\mathbf{x}}-\mathbf{o}\right)
.\label{f11}%
\end{equation}
One notes the special case,%
\begin{equation}
\mathsf{Q}=\mathsf{I}\text{ and }\mathbf{o}=\mathbf{v}_{0}t+\mathbf{x}%
_{0},\label{f10.5}%
\end{equation}
where $\mathbf{v}_{0}$\ and $\mathbf{x}_{0}$\ are arbitrary constant vectors,
is a Galilean transformation. Taking the time derivative of (\ref{f9}) gives
the velocity of the moving frame,%
\begin{equation}
\widehat{\mathbf{v}}_{f}=\frac{d\mathsf{Q}}{dt}\cdot\mathbf{x}+\frac
{d\mathbf{o}}{dt},\label{f12}%
\end{equation}
or upon substitution of (\ref{f11}),%
\begin{equation}
\widehat{\mathbf{v}}_{f}=\frac{d\mathsf{Q}}{dt}\cdot\mathsf{Q}^{T}\cdot\left(
\widehat{\mathbf{x}}-\mathbf{o}\right)  +\frac{d\mathbf{o}}{dt}.\label{f13}%
\end{equation}

Next, let us consider how fluid properties are transformed into the moving
reference frame.\ Suppose $A\left(  \mathbf{x},t\right)  $ is an order $n$
tensor representing a local variable in the fixed frame. Then, in general, the
$i_{1}i_{2}\ldots i_{n}$-component of $A$\ is transformed into the moving
coordinate system as%
\begin{equation}
\widehat{A}_{i_{1}i_{2}\ldots i_{n}}=\sum\limits_{j_{1}=1}^{3}\sum
\limits_{j_{2}=1}^{3}\ldots\sum\limits_{j_{n}=1}^{3}Q_{i_{1}j_{1}}%
Q_{i_{2}j_{2}}\cdots Q_{i_{n}j_{n}}A_{j_{1}j_{2}\ldots j_{n}}+\left(
\widehat{A}_{f}\right)  _{i_{1}i_{2}\ldots i_{n}},\label{f14}%
\end{equation}
where the sum is the mathematical transformation of $A$ into the moving
reference frame when rotation is involved and the last term is the possible
contribution to $\widehat{A}$ from the moving frame itself. For example, when
$A$ represents a\ scalar, $\alpha$, a vector, $\mathbf{w}$, or an order $2$
tensor, $\mathsf{M}$, then (\ref{f14}) becomes%
\begin{equation}
\widehat{\alpha}=\alpha+\widehat{\alpha}_{f},\label{f15}%
\end{equation}%
\begin{equation}
\widehat{\mathbf{w}}=\mathsf{Q}\cdot\mathbf{w}+\widehat{\mathbf{w}}%
_{f},\label{f16}%
\end{equation}
and%
\begin{equation}
\widehat{\mathsf{M}}=\mathsf{Q}\cdot\mathsf{M}\cdot\mathsf{Q}^{T}+\widehat
{M}_{f},\label{f17}%
\end{equation}
respectively.

Any property $A$\ is called objective if $\widehat{A}_{f}=0$, or in other
words, if it has no additional contribution due to the movement of its
reference frame. Local thermodynamic variables -- such as mass density,
internal energy density, entropy density, thermodynamic pressure, and
temperature -- are all regarded as objective, e.g. the mass density and
temperature of a fluid are expected to remain unaffected in a moving reference
frame:%
\begin{equation}
\widehat{\rho}=\rho\text{ and }\widehat{T}=T.\label{f19}%
\end{equation}
Transport coefficients, such as viscosity and thermal conductivity, which can
have functional dependence on the thermodynamic variables, are also expected
to be objective. It is generally a straight-forward matter to consider
mathematical operations on objective quantities to see whether or not the
results are objective. For example, property (\ref{p1}) implies the gradient
of any objective order $n$\ tensor is an objective order $n+1$ tensor, and in
particular, relations (\ref{f19}) and the chain rule, lead to%
\begin{equation}
\widehat{\nabla\rho}=\mathsf{Q}\cdot\nabla\rho\text{ and }\widehat{\nabla
T}=\mathsf{Q}\cdot\nabla T.\label{p1.05}%
\end{equation}

On the other hand, when $A$\ represents a mechanical variable, there may
sometimes arise a non-zero moving frame contribution, $\widehat{A}_{f}$. Such
is the case for the local fluid velocity, $\mathbf{v}$. Employing vector
transformation equation (\ref{f16}) and expression (\ref{f13}) for
$\widehat{\mathbf{v}}_{f}$, we find the fluid velocity in the moving reference
frame is given by%
\begin{equation}
\widehat{\mathbf{v}}=\mathsf{Q}\cdot\mathbf{v}+\frac{d\mathsf{Q}}{dt}%
\cdot\mathsf{Q}^{T}\cdot\left(  \widehat{\mathbf{x}}-\mathbf{o}\right)
+\frac{d\mathbf{o}}{dt}.\label{f20}%
\end{equation}
As computations \ref{com2}-\ref{com4} in Appendix \ref{com}, we use the above
equation together with a few tensor identities and the chain rule to compute
the following expressions for the velocity gradient, the divergence of the
velocity, and the symmetric deviatoric part of the velocity gradient in the
moving reference frame:%
\begin{equation}
\widehat{\nabla\mathbf{v}}=\mathsf{Q}\cdot\nabla\mathbf{v}\cdot\mathsf{Q}%
^{T}+\mathsf{Q}\cdot\frac{d\mathsf{Q}^{T}}{dt},\label{f21}%
\end{equation}%
\begin{equation}
\widehat{\nabla\cdot\mathbf{v}}=\nabla\cdot\mathbf{v},\label{f22}%
\end{equation}
and%
\begin{equation}
\widehat{\nabla\mathbf{v}^{SD}}=\mathsf{Q}\cdot\nabla\mathbf{v}^{SD}%
\cdot\mathsf{Q}^{T},\label{f23}%
\end{equation}
respectively. As we can see, the quantities, $\nabla\cdot\mathbf{v}$\ and
$\nabla\mathbf{v}^{SD}$,\ are objective; however, $\nabla\mathbf{v}$\ is not.

\subsection{\label{vst}The viscous stress tensor and dcE analog}

From Equation (\ref{a6}), we compute the NSF viscous stress tensor in the
moving reference frame as%
\begin{align*}
\widehat{\mathsf{S}}  & =\widehat{\zeta}\left(  \widehat{\nabla\cdot
\mathbf{v}}\right)  \mathsf{I}+2\widehat{\eta}\widehat{\nabla\mathbf{v}^{SD}%
}\\
& =\zeta\left(  \nabla\cdot\mathbf{v}\right)  \mathsf{I}+2\eta\mathsf{Q}%
\cdot\nabla\mathbf{v}^{SD}\cdot\mathsf{Q}^{T}\\
& =\mathsf{Q}\cdot\left(  \zeta\left(  \nabla\cdot\mathbf{v}\right)
\mathsf{I}+2\eta\nabla\mathbf{v}^{SD}\right)  \cdot\mathsf{Q}^{T}\\
& =\mathsf{Q}\cdot\mathsf{S}\cdot\mathsf{Q}^{T},
\end{align*}
where we have used (\ref{f22}), (\ref{f23}) and the objectivity of $\zeta
$\ and $\eta$ in the second line and the orthogonality of $\mathsf{Q}$\ in the
third line. Thus, we have shown the NSF viscous stress tensor to be objective.

Next, from Equation (\ref{f7.5}) and the product rule, one computes the
following expression for the dcE model's analog viscous stress tensor:%
\begin{equation}
\mathsf{S}_{dcE}=\nu\left(  \rho\nabla\mathbf{v}+\left(  \nabla\rho\right)
\mathbf{v}\right)  .\label{v3}%
\end{equation}
In the moving reference frame, the above is%
\begin{align*}
\widehat{\mathsf{S}_{dcE}}  & =\widehat{\nu}\left(  \widehat{\rho}%
\widehat{\nabla\mathbf{v}}+\left(  \widehat{\nabla\rho}\right)  \widehat
{\mathbf{v}}\right)  \\
& =\nu\left(
\begin{array}
[c]{c}%
\rho\left(  \mathsf{Q}\cdot\nabla\mathbf{v}\cdot\mathsf{Q}^{T}+\mathsf{Q}%
\cdot\frac{d\mathsf{Q}^{T}}{dt}\right)  +\\
\mathsf{Q}\cdot\left(  \nabla\rho\right)  \left(  \mathsf{Q}\cdot
\mathbf{v}+\frac{d\mathsf{Q}}{dt}\cdot\mathsf{Q}^{T}\cdot\left(
\widehat{\mathbf{x}}-\mathbf{o}\right)  +\frac{d\mathbf{o}}{dt}\right)
\end{array}
\right)  \\
& =\mathsf{Q}\cdot\mathsf{S}_{dcE}\cdot\mathsf{Q}^{T}+\left(  \widehat
{\mathsf{S}_{dcE}}\right)  _{f},
\end{align*}
where the frame-dependent term is given by%
\begin{equation}
\left(  \widehat{\mathsf{S}_{dcE}}\right)  _{f}=\nu\left(  \rho\mathsf{Q}%
\cdot\frac{d\mathsf{Q}^{T}}{dt}+\mathsf{Q}\cdot\left(  \nabla\rho\right)
\left(  \frac{d\mathsf{Q}}{dt}\cdot\mathsf{Q}^{T}\cdot\left(  \widehat
{\mathbf{x}}-\mathbf{o}\right)  +\frac{d\mathbf{o}}{dt}\right)  \right)
.\label{v6}%
\end{equation}
The objectivity of $\rho$ and $\nu$ and Equations (\ref{p1.05}) and
(\ref{f21}) have been used in the second line of the above computation, and
Equations (\ref{v3}) and (\ref{f17}) have been used to obtain the final
result. We see that $\left(  \widehat{\mathsf{S}_{dcE}}\right)  _{f}%
$\ vanishes only when the rotation, $\mathsf{Q}$,\ and the origin,
$\mathbf{o}$, of the $\widehat{\mathbf{x}}$ reference frame are constant
quantities that do not depend on time -- or, in other words, when the
$\widehat{\mathbf{x}}$ reference frame is not moving.\ Consequently, the
analog viscous stress tensor, $\mathsf{S}_{dcE}$, in the dcE model\ is not an
objective order $2$\ tensor. In fact, by substituting choices (\ref{f10.5})
into the above, one finds%
\begin{equation}
\left(  \widehat{\mathsf{S}_{dcE}}\right)  _{f}=\nu\left(  \nabla\rho\right)
\mathbf{v}_{0},\label{v6.5}%
\end{equation}
which demonstrates that $\mathsf{S}_{dcE}$ is not even Galilean
invariant.\footnote{We wish to clarify that, as remarked in \textsc{Sv\"{a}rd}
\cite[Sec. 5.3]{sv1}, the dcE model's momentum equation (\ref{f4}) satisfies
the principle of Galilean invariance, whereas Equation (\ref{v6.5}) implies
the constitutive law appearing in that equation is not Galilean invariant.}

Now, we are equipped to identify the fallacy behind the following statement in
\textsc{Sv\"{a}rd} \cite[Sec. 5]{sre}:

\begin{quote}
"...if the actual constitutive laws are considered, i.e., the diffusive
fluxes, they all take the same form as Fourier's law. Hence, and contrary to
Morris' claim, \textit{the constitutive laws of (1)} [the dcE model]
\textit{are indeed frame indifferent}..."\footnote{The italics for emphasis
were introduced in \textsc{Sv\"{a}rd} \cite{sre} and not by myself.}
\end{quote}

As we pointed out in Section \ref{et}, the gradient of an objective quantity
is another objective quantity of one increased tensorial order. Therefore,
since the temperature is an objective scalar, Fourier's law (\ref{a5}) yields
a heat flux that is, indeed, an objective vector satisfying reference frame
indifference. By a similar argument, it follows that the dcE model's mass
diffusion flux, $-\nu\nabla\rho$, and non-convective internal energy flux,
$-\nu\nabla u-\kappa_{T}\nabla T$, are also independent of the moving
reference frame. The momentum density, $\rho\mathbf{v}$, however, is not an
objective quantity and so one may not use the aforementioned property to
reason that its gradient is frame indifferent. Only by carrying out the
computation of $\nabla\left(  \rho\mathbf{v}\right)  $ under the Euclidean
transformation (\ref{f9}), as we have done above, may we learn whether or not
it is an objective order $2$\ tensor, and we see that it is not.

\subsection{\label{vd}The viscous dissipation and dcE analog}

From Equation (\ref{f2}), we compute the NSF viscous dissipation in the moving
reference frame as%
\begin{align*}
\widehat{\Psi}  & =\widehat{\zeta}\left(  \widehat{\nabla\cdot\mathbf{v}%
}\right)  ^{2}+2\widehat{\eta}\widehat{\nabla\mathbf{v}^{SD}}\cdot
\cdot\widehat{\nabla\mathbf{v}^{SD}}\\
& =\zeta\left(  \nabla\cdot\mathbf{v}\right)  ^{2}+2\eta\left(  \mathsf{Q}%
\cdot\nabla\mathbf{v}^{SD}\cdot\mathsf{Q}^{T}\right)  \cdot\cdot\left(
\mathsf{Q}\cdot\nabla\mathbf{v}^{SD}\cdot\mathsf{Q}^{T}\right) \\
& =\zeta\left(  \nabla\cdot\mathbf{v}\right)  ^{2}+2\eta\nabla\mathbf{v}%
^{SD}\cdot\cdot\nabla\mathbf{v}^{SD}\\
& =\Psi,
\end{align*}
where (\ref{f22}), (\ref{f23}) and the objectivity of $\zeta$\ and $\eta
$\ have been used in the second line, and property (\ref{p2}) has been used in
the third line. This shows the NSF viscous dissipation term, $\Psi$, to be an
objective quantity.

On the other hand, computing the dcE model's analog viscous dissipation term
in the moving reference frame from Equation (\ref{f8}) yields%
\begin{align*}
\widehat{\Psi_{dcE}}  & =\widehat{\rho}\widehat{\nu}\widehat{\nabla\mathbf{v}%
}\cdot\cdot\widehat{\nabla\mathbf{v}}\\
& =\rho\nu\left(  \mathsf{Q}\cdot\nabla\mathbf{v}\cdot\mathsf{Q}%
^{T}+\mathsf{Q}\cdot\frac{d\mathsf{Q}^{T}}{dt}\right)  \cdot\cdot\left(
\mathsf{Q}\cdot\nabla\mathbf{v}\cdot\mathsf{Q}^{T}+\mathsf{Q}\cdot
\frac{d\mathsf{Q}^{T}}{dt}\right) \\
& =\rho\nu\left(
\begin{array}
[c]{c}%
\left(  \mathsf{Q}\cdot\nabla\mathbf{v}\cdot\mathsf{Q}^{T}\right)  \cdot
\cdot\left(  \mathsf{Q}\cdot\nabla\mathbf{v}\cdot\mathsf{Q}^{T}\right)  +\\
\left(  \mathsf{Q}\cdot\nabla\mathbf{v}\cdot\mathsf{Q}^{T}\right)  \cdot
\cdot\left(  \mathsf{Q}\cdot\frac{d\mathsf{Q}^{T}}{dt}\right)  +\left(
\mathsf{Q}\cdot\frac{d\mathsf{Q}^{T}}{dt}\right)  \cdot\cdot\left(
\mathsf{Q}\cdot\nabla\mathbf{v}\cdot\mathsf{Q}^{T}\right)  +\\
\left(  \mathsf{Q}\cdot\frac{d\mathsf{Q}^{T}}{dt}\right)  \cdot\cdot\left(
\mathsf{Q}\cdot\frac{d\mathsf{Q}^{T}}{dt}\right)
\end{array}
\right) \\
& =\rho\nu\left(  \nabla\mathbf{v}\cdot\cdot\nabla\mathbf{v}+2\nabla
\mathbf{v}\cdot\cdot\left(  \frac{d\mathsf{Q}}{dt}\cdot\mathsf{Q}^{T}\right)
+\frac{d\mathsf{Q}}{dt}\cdot\cdot\frac{d\mathsf{Q}}{dt}\right) \\
& =\Psi_{dcE}+\left(  \widehat{\Psi_{dcE}}\right)  _{f},
\end{align*}
where the frame-dependent contribution is given by%
\begin{equation}
\left(  \widehat{\Psi_{dcE}}\right)  _{f}=\rho\nu\left(  2\nabla
\mathbf{v}\cdot\cdot\left(  \frac{d\mathsf{Q}^{T}}{dt}\cdot\mathsf{Q}\right)
+\frac{d\mathsf{Q}}{dt}\cdot\cdot\frac{d\mathsf{Q}}{dt}\right)  .\label{f27}%
\end{equation}
In the computation of $\widehat{\Psi_{dcE}}$\ above, Equation (\ref{f21}) and
the objectivity of $\rho$\ and $\nu$\ have been used in the second line, and
properties (\ref{p1.1}), (\ref{p2}), and (\ref{p3.1}) have been used in the
fourth line. We see that $\left(  \widehat{\Psi_{dcE}}\right)  _{f}$\ vanishes
only when the rotation, $\mathsf{Q}$,\ of the reference frame is a constant
tensor -- or, in other words, when the $\widehat{\mathbf{x}}$ reference frame
has no rotational motion -- and therefore the analog viscous dissipation term,
$\Psi_{dcE}$, in the dcE model\ is not an objective scalar.

\subsection{\label{vrep}The volumetric rate of entropy production}

Using the results of the previous section, it is now trivial to show that the
volumetric rate of entropy production in the NSF model is objective. In the
moving reference frame, Equation (\ref{ff2}) yields%
\begin{align*}
\widehat{\sigma}  & =\frac{\widehat{\lambda}}{\widehat{T}^{2}}\widehat{\nabla
T}\cdot\widehat{\nabla T}+\frac{\widehat{\Psi}}{\widehat{T}}\\
& =\frac{\lambda}{T^{2}}\left(  \mathsf{Q}\cdot\nabla T\right)  \cdot\left(
\mathsf{Q}\cdot\nabla T\right)  +\frac{\Psi}{T}\\
& =\sigma,
\end{align*}
where (\ref{p1.05}) and the objectivity of $T$, $\lambda$, and $\Psi$ have
been used in the second line, and identity (\ref{p.1}) has been used in the
final line.

In a similar manner, we compute the dcE model's volumetric entropy production
rate in the moving reference frame from Equation (\ref{ff5}) as%
\begin{align*}
\widehat{\sigma_{dcE}}  & =\frac{1}{\widehat{T}^{2}}\left(  \frac
{\widehat{\rho}\widehat{c_{p}}}{\widehat{\gamma}}\widehat{\nu}+\widehat
{\kappa_{T}}\right)  \widehat{\nabla T}\cdot\widehat{\nabla T}+\frac
{\widehat{c}^{2}\widehat{\nu}}{\widehat{\rho}\widehat{T}\widehat{\gamma}%
}\widehat{\nabla\rho}\cdot\widehat{\nabla\rho}+\frac{\widehat{\Psi_{dcE}}%
}{\widehat{T}}\\
& =\left(
\begin{array}
[c]{c}%
\frac{1}{T^{2}}\left(  \frac{\rho c_{p}}{\gamma}\nu+\kappa_{T}\right)  \left(
\mathsf{Q}\cdot\nabla T\right)  \cdot\left(  \mathsf{Q}\cdot\nabla T\right)
+\\
\frac{c^{2}\nu}{\rho T\gamma}\left(  \mathsf{Q}\cdot\nabla\rho\right)
\cdot\left(  \mathsf{Q}\cdot\nabla\rho\right)  +\frac{\Psi_{dcE}}{T}%
+\frac{\left(  \widehat{\Psi_{dcE}}\right)  _{f}}{T}%
\end{array}
\right) \\
& =\sigma_{dcE}+\left(  \widehat{\sigma_{dcE}}\right)  _{f}%
\end{align*}
with frame-dependent contribution given by%
\[
\left(  \widehat{\sigma_{dcE}}\right)  _{f}=\frac{\left(  \widehat{\Psi_{dcE}%
}\right)  _{f}}{T},
\]
or using Equation (\ref{f27}) from the previous section,%
\begin{equation}
\left(  \widehat{\sigma_{dcE}}\right)  _{f}=\frac{\rho\nu}{T}\left(
2\nabla\mathbf{v}\cdot\cdot\left(  \frac{d\mathsf{Q}^{T}}{dt}\cdot
\mathsf{Q}\right)  +\frac{d\mathsf{Q}}{dt}\cdot\cdot\frac{d\mathsf{Q}}%
{dt}\right)  .\label{ff12}%
\end{equation}
As discussed in Section \ref{intro}, the volumetric rate of entropy production
is physically required to be an objective scalar. Furthermore, even if one
were to disagree with the expectation that $\Psi_{dcE}$\ and $\sigma_{dcE}$
satisfy objectivity, then any attempt to justify the moving frame-dependence
is additionally hampered by the fact that in Equation (\ref{ff12}) the first
term on the right-hand side,%
\[
\frac{2\rho\nu}{T}\nabla\mathbf{v}\cdot\cdot\left(  \frac{d\mathsf{Q}^{T}}%
{dt}\cdot\mathsf{Q}\right)  ,
\]
is not, in general, positive semi-definite and thereby can lead to a violation
of the second law of thermodynamics.

Let us consider, for example, a reference frame that rotates steadily about
the $x_{3}$-axis. If the rotation of the frame occurs in the $x_{1}$ to
$x_{2}$\ direction at a constant angular frequency of $\omega_{f}$, then the
coordinates in the moving system are given by%
\begin{equation}
\widehat{\mathbf{x}}=\mathsf{Q}\left(  t\right)  \cdot\mathbf{x}\label{ff13}%
\end{equation}
with the following rotation tensor:%
\begin{equation}
\mathsf{Q}=\left[
\begin{array}
[c]{ccc}%
\cos\left(  \omega_{f}t\right)  & \sin\left(  \omega_{f}t\right)  & 0\\
-\sin\left(  \omega_{f}t\right)  & \cos\left(  \omega_{f}t\right)  & 0\\
0 & 0 & 1
\end{array}
\right]  .\label{ff14}%
\end{equation}
Using the above in Equation (\ref{ff12}) to compute the frame-dependent
contribution to the dcE model's volumetric entropy production rate, one finds%
\begin{equation}
\left(  \widehat{\sigma_{dcE}}\right)  _{f}=\frac{2\rho\nu}{T}\omega
_{f}\left(  \frac{\partial v_{1}}{\partial x_{2}}-\frac{\partial v_{2}%
}{\partial x_{1}}+\omega_{f}\right)  ,\label{ff15}%
\end{equation}
which is not positive semi-definite unless the velocity of the flow is
everywhere constrained to satisfy%
\begin{equation}
\frac{\partial v_{1}}{\partial x_{2}}=\frac{\partial v_{2}}{\partial x_{1}%
},\label{ff16}%
\end{equation}
i.e., to be irrotational about the $x_{3}$-axis. Even when the above condition
is met, we fail to see any physical justification for the dcE model's
predicted frame-dependent increase to its volumetric entropy production rate:%
\begin{equation}
\left(  \widehat{\sigma_{dcE}}\right)  _{f}=\frac{2\rho\nu}{T}\omega_{f}%
^{2}.\label{ff17}%
\end{equation}

\section{Summary and conclusions}

We have conducted a study into the implications of the dcE model's analog
viscous stress tensor,%
\[
\mathsf{S}_{dcE}=\nu\nabla\left(  \rho\mathbf{v}\right)  .
\]
As part of this study, we have derived the dcE model's balance laws for the
kinetic and internal energy in Section \ref{sem} and for the entropy in
Appendix \ref{apps}, and we have compared them to the corresponding equations
in the traditional NSF model. Doing so has allowed us, for the dcE model, to
identify its analog viscous dissipation term:%
\[
\Psi_{dcE}=\rho\nu\nabla\mathbf{v}\cdot\cdot\nabla\mathbf{v},
\]
and to determine its volumetric entropy production rate:%
\[
\sigma_{dcE}=\frac{1}{T^{2}}\left(  \frac{\rho c_{p}\nu}{\gamma}+\kappa
_{T}\right)  \nabla T\cdot\nabla T+\frac{c^{2}\nu}{\rho T\gamma}\nabla
\rho\cdot\nabla\rho+\frac{\Psi_{dcE}}{T}%
\]
and the resulting restrictions on its diffusion coefficients from the second
law of thermodynamics:%
\[
\nu\geq0\text{ and }\kappa_{T}\geq-\frac{\rho c_{p}}{\gamma}\nu.
\]

Throughout our computations in Section \ref{obj}, we have presented the
mathematics behind determining whether or not\ a local Eulerian quantity is
objective, thereby circumventing the Lagrangian description that is the basis
for the principle of material frame indifference. By performing a general
Euclidean transformation on the dcE analogs of the viscous stress tensor and
viscous dissipation and the dcE model's volumetric rate of entropy production,
we have shown that, unlike their NSF counterparts, the foregoing quantities
are not objective and their frame-dependent terms are given by%
\[
\left(  \widehat{\mathsf{S}_{dcE}}\right)  _{f}=\nu\left(  \rho\mathsf{Q}%
\cdot\frac{d\mathsf{Q}^{T}}{dt}+\mathsf{Q}\cdot\left(  \nabla\rho\right)
\left(  \frac{d\mathsf{Q}}{dt}\cdot\mathsf{Q}^{T}\cdot\left(  \widehat
{\mathbf{x}}-\mathbf{o}\right)  +\frac{d\mathbf{o}}{dt}\right)  \right)  ,
\]%
\[
\left(  \widehat{\Psi_{dcE}}\right)  _{f}=\rho\nu\left(  2\nabla
\mathbf{v}\cdot\cdot\left(  \frac{d\mathsf{Q}^{T}}{dt}\cdot\mathsf{Q}\right)
+\frac{d\mathsf{Q}}{dt}\cdot\cdot\frac{d\mathsf{Q}}{dt}\right)  ,
\]
and%
\[
\left(  \widehat{\sigma_{dcE}}\right)  _{f}=\frac{\rho\nu}{T}\left(
2\nabla\mathbf{v}\cdot\cdot\left(  \frac{d\mathsf{Q}^{T}}{dt}\cdot
\mathsf{Q}\right)  +\frac{d\mathsf{Q}}{dt}\cdot\cdot\frac{d\mathsf{Q}}%
{dt}\right)  ,
\]
respectively. Furthermore, we have demonstrated in the special case of a
Galilean transformation that there is a frame-dependent term for the dcE
analog viscous stress tensor given by%
\[
\left(  \widehat{\mathsf{S}_{dcE}}\right)  _{f}=\nu\left(  \nabla\rho\right)
\mathbf{v}_{0},
\]
meaning that the constitutive equation for $\mathsf{S}_{dcE}$\ is not Galilean invariant.

If one disagrees with the assumptions that $\mathsf{S}_{dcE}$, $\Psi_{dcE}$,
and $\sigma_{dcE}$ should be objective -- and in the case of $\mathsf{S}%
_{dcE}$, Galilean invariant -- then one is challenged to provide physical
arguments for the presence of the above terms. However, we argue that the
volumetric entropy production rate is physically required to be an objective
scalar so that reference frame motion cannot contribute to a fluid's
dissipation; and we further point out that the above frame-dependent
contribution, $\left(  \widehat{\sigma_{dcE}}\right)  _{f}$, can potentially
violate the second law of thermodynamics. From this, we conclude that when
time-dependent rotations are allowed (even those that are steady), the dcE
model exhibits an unphysical dependence on moving reference frames.

\appendix{}

\section{\label{apps}Entropy balance and the second law of thermodynamics}

Following \textsc{Callen} \cite[Ch. 1.9]{callen}, let us propose a fundamental
equation for the entropy density of a single-component fluid of the form,%
\begin{equation}
s_{eq}=s_{\left(  s\right)  }\left(  u_{eq},\rho_{eq}\right)  ,\label{ce1}%
\end{equation}
where the subscript "$eq$" is used in this appendix to indicate that the
quantity corresponds to a uniform, macroscopic system in thermodynamic
equilibrium. Next, we note a few useful equilibrium thermodynamic
relationships. The intensive parameters are defined via the partial
derivatives of (\ref{ce1}):%
\begin{equation}
\frac{1}{T_{eq}}=\frac{\partial s_{\left(  s\right)  }}{\partial u_{eq}}\text{
and }\frac{\mu_{eq}}{T_{eq}}=-\frac{\partial s_{\left(  s\right)  }}%
{\partial\rho_{eq}},\label{ce2}%
\end{equation}
where the symbol $\mu$\ is used to represent the chemical potential, and with
these the first differential of (\ref{ce1}) can be expressed as%
\begin{equation}
ds_{eq}=\frac{1}{T_{eq}}du_{eq}-\frac{\mu_{eq}}{T_{eq}}d\rho_{eq}%
.\label{ce2.15}%
\end{equation}
Also from \textsc{Callen} \cite[Ch. 3.1 and 3.2]{callen}, we have the Euler
relation,%
\begin{equation}
u_{eq}=T_{eq}s_{eq}-p_{eq}+\mu_{eq}\rho_{eq},\label{ce2.1}%
\end{equation}
and the Gibbs-Duhem relation, which can be expressed as%
\begin{equation}
d\left(  \frac{p_{eq}}{T_{eq}}\right)  =-u_{eq}d\left(  \frac{1}{T_{eq}%
}\right)  +\rho_{eq}d\left(  \frac{\mu_{eq}}{T_{eq}}\right)  .\label{ce2.2}%
\end{equation}
Lastly, by using the Legendre transformation for the Helmholtz potential as
described in \textsc{Callen} \cite[Ch. 5.3]{callen}, along with the standard
thermodynamic definitions of the adiabatic sound speed, $c$, the coefficient
of thermal expansion, $\alpha_{p}$, the isobaric specific heat, $c_{p}$, and
the ratio of isobaric to isochoric specific heat, $\gamma$, one may derive the
following differential relationships:%
\begin{equation}
du_{eq}=\frac{\rho_{eq}\left(  c_{p}\right)  _{eq}}{\gamma_{eq}}%
dT_{eq}+\left(  \frac{u_{eq}+p_{eq}}{\rho_{eq}}-\frac{T_{eq}\left(  \alpha
_{p}\right)  _{eq}c_{eq}^{2}}{\gamma_{eq}}\right)  d\rho_{eq}\label{ce2.3}%
\end{equation}
and%
\begin{equation}
d\left(  \frac{\mu_{eq}}{T_{eq}}\right)  =-\frac{1}{T_{eq}^{2}}\left(
\frac{u_{eq}+p_{eq}}{\rho_{eq}}-\frac{T_{eq}\left(  \alpha_{p}\right)
_{eq}c_{eq}^{2}}{\gamma_{eq}}\right)  dT_{eq}+\frac{c_{eq}^{2}}{\rho
_{eq}T_{eq}\gamma_{eq}}d\rho_{eq}.\label{ce2.4}%
\end{equation}

Assuming our phenomena of study do not stray too far from equilibrium, we may
apply the standard assumptions and techniques of irreversible thermodynamics
that are detailed in \textsc{de Groot} and \textsc{Mazur} \cite[Ch. III and
IV]{degroot}, for example. First, let us employ the local equilibrium
hypothesis in which one assumes that if $a_{eq}$\ represents any of the
thermodynamic parameters mentioned above, then its corresponding local
variable is given by%
\begin{equation}
a=\left.  a_{eq}\right\vert _{\left(  \mathbf{x},t\right)  }.\label{ce5}%
\end{equation}
In particular, the fundamental equation (\ref{ce1}) is assumed to hold locally
as%
\begin{equation}
s=s_{\left(  s\right)  }\left(  u,\rho\right)  ,\label{ce6}%
\end{equation}
and using the local version of equation (\ref{ce2.15}), its partial time
derivative is computed to be%
\begin{equation}
\frac{\partial s}{\partial t}=\frac{1}{T}\frac{\partial u}{\partial t}%
-\frac{\mu}{T}\frac{\partial\rho}{\partial t}.\label{ce7}%
\end{equation}
The entropy balance law takes the following general form:%
\begin{equation}
\frac{\partial s}{\partial t}=-\nabla\cdot\left(  \mathbf{q}_{S}%
+s\mathbf{v}\right)  +\sigma,\label{ce7.3}%
\end{equation}
where $\mathbf{q}_{S}$\ is the non-convective entropy flux and $\sigma$\ is
the volumetric rate of entropy production. Next, we may substitute the above,
along with mass and internal energy balance laws for the NSF and dcE models,
into relation (\ref{ce7})\ in order to derive expressions for $\mathbf{q}_{S}%
$\ and $\sigma$. Doing so, with NSF equations (\ref{a1}) and (\ref{f1}) yields%
\[
-\nabla\cdot\left(  \mathbf{q}_{S}+s\mathbf{v}\right)  +\sigma=-\frac{1}%
{T}\nabla\cdot\left(  \mathbf{q}+u\mathbf{v}\right)  -\frac{p}{T}\nabla
\cdot\mathbf{v}+\frac{\Psi}{T}+\frac{\mu}{T}\nabla\cdot\left(  \rho
\mathbf{v}\right)
\]
or using the product rule,%
\begin{multline*}
-\nabla\cdot\left(  \mathbf{q}_{S}+s\mathbf{v}\right)  +\sigma=-\nabla
\cdot\left(  \frac{\mathbf{q}}{T}+\frac{1}{T}\left(  u+p-\frac{\rho\mu}%
{T}\right)  \mathbf{v}\right) \\
+\mathbf{v}\cdot\left(  u\nabla\left(  \frac{1}{T}\right)  +\nabla\left(
\frac{p}{T}\right)  -\rho\nabla\left(  \frac{\mu}{T}\right)  \right)
+\mathbf{q}\cdot\nabla\left(  \frac{1}{T}\right)  +\frac{\Psi}{T}.
\end{multline*}
Note that the local forms of Euler relation (\ref{ce2.1}) and Gibbs-Duhem
relation (\ref{ce2.2}) imply, respectively,%
\begin{equation}
s=\frac{1}{T}\left(  u+p-\frac{\rho\mu}{T}\right) \label{cc2}%
\end{equation}
and%
\begin{equation}
\nabla\left(  \frac{p}{T}\right)  =-u\nabla\left(  \frac{1}{T}\right)
+\rho\nabla\left(  \frac{\mu}{T}\right)  ,\label{cc3}%
\end{equation}
and these, when substituted into the above, yield the following expression for
$\sigma$:%
\begin{equation}
\sigma=-\nabla\cdot\left(  \frac{\mathbf{q}}{T}-\mathbf{q}_{S}\right)
+\nabla\left(  \frac{1}{T}\right)  \cdot\mathbf{q}+\frac{\Psi}{T},\label{cc4}%
\end{equation}
or with Fourier's law (\ref{a5}) and Equation (\ref{f2}) for the NSF viscous
dissipation,%
\begin{equation}
\sigma=-\nabla\cdot\left(  -\frac{\lambda}{T}\nabla T-\mathbf{q}_{S}\right)
+\frac{\lambda}{T^{2}}\nabla T\cdot\nabla T+\frac{\zeta}{T}\left(  \nabla
\cdot\mathbf{v}\right)  ^{2}+\frac{2\eta}{T}\nabla\mathbf{v}^{SD}\cdot
\cdot\nabla\mathbf{v}^{SD}.\label{cc5}%
\end{equation}
The second law of thermodynamics requires that $\sigma$ is non-negative, and
we see from Equation (\ref{cc5}) that this condition is guaranteed to be
upheld if the non-convective entropy flux is given by%
\begin{equation}
\mathbf{q}_{S}=-\frac{\lambda}{T}\nabla T\label{cc6}%
\end{equation}
and if the thermal conductivity, $\lambda$, and the bulk and shear
viscosities, $\zeta$\ and $\eta$, are each non-negative.

In a similar manner, one may substitute (\ref{ce7.3}) and dcE mass and
internal energy balance laws (\ref{f3}) and (\ref{f7}) into relation
(\ref{ce7}) to find%
\[
-\nabla\cdot\left(  \mathbf{q}_{S}+s\mathbf{v}\right)  +\sigma=-\frac{1}%
{T}\nabla\cdot\left(  -\nu\nabla u-\kappa_{T}\nabla T+u\mathbf{v}\right)
-\frac{p}{T}\nabla\cdot\mathbf{v}+\frac{\rho\nu}{T}\nabla\mathbf{v}\cdot
\cdot\nabla\mathbf{v}+\frac{\mu}{T}\nabla\cdot\left(  -\nu\nabla\rho
+\rho\mathbf{v}\right)  ,
\]
or using the product rule,%
\begin{multline*}
-\nabla\cdot\left(  \mathbf{q}_{S}+s\mathbf{v}\right)  +\sigma=-\nabla
\cdot\left(  -\frac{\nu}{T}\left(  \nabla u-\mu\nabla\rho\right)
-\frac{\kappa_{T}}{T}\nabla T+\frac{1}{T}\left(  u+p-\frac{\rho\mu}{T}\right)
\mathbf{v}\right) \\
+\mathbf{v}\cdot\left(  u\nabla\left(  \frac{1}{T}\right)  +\nabla\left(
\frac{p}{T}\right)  -\rho\nabla\left(  \frac{\mu}{T}\right)  \right) \\
-\left(  \nu\nabla u+\kappa_{T}\nabla T\right)  \cdot\nabla\left(  \frac{1}%
{T}\right)  +\nu\nabla\rho\cdot\nabla\left(  \frac{\mu}{T}\right)  +\frac
{\rho\nu}{T}\nabla\mathbf{v}\cdot\cdot\nabla\mathbf{v}.
\end{multline*}
If one employs the following local forms of equilibrium thermodynamic
relations (\ref{ce2.15}), (\ref{ce2.3}), and (\ref{ce2.4}):%
\[
\nabla s=\frac{1}{T}\nabla u-\frac{\mu}{T}\nabla\rho,
\]%
\[
\nabla u=\frac{\rho c_{p}}{\gamma}\nabla T+\left(  \frac{u+p}{\rho}%
-\frac{T\alpha_{p}c^{2}}{\gamma}\right)  \nabla\rho,
\]
and%
\[
\nabla\left(  \frac{\mu}{T}\right)  =-\frac{1}{T^{2}}\left(  \frac{u+p}{\rho
}-\frac{T\alpha_{p}c^{2}}{\gamma}\right)  \nabla T+\frac{c^{2}}{\rho T\gamma
}\nabla\rho,
\]
and, again, uses local Euler and Gibbs-Duhem relations (\ref{cc2}) and
(\ref{cc3}), then from the above equation one obtains the following expression
for the volumetric rate of entropy production:%
\[
\sigma=-\nabla\cdot\left(  -\nu\nabla s-\frac{\kappa_{T}}{T}\nabla
T-\mathbf{q}_{S}\right)  +\frac{1}{T^{2}}\left(  \frac{\rho c_{p}\nu}{\gamma
}+\kappa_{T}\right)  \nabla T\cdot\nabla T+\frac{c^{2}\nu}{\rho T\gamma}%
\nabla\rho\cdot\nabla\rho+\frac{\rho\nu}{T}\nabla\mathbf{v}\cdot\cdot
\nabla\mathbf{v}.
\]
Therefore, we see that the second law of thermodynamics holds in the dcE model
if the non-convective entropy flux is given by%
\begin{equation}
\mathbf{q}_{S}=-\nu\nabla s-\frac{\kappa_{T}}{T}\nabla T\label{cc8}%
\end{equation}
and if the dcE diffusion coefficients satisfy%
\begin{equation}
\nu\geq0\text{ and }\kappa_{T}\geq-\frac{\rho c_{p}}{\gamma}\nu.\label{cc9}%
\end{equation}

\section{\label{ten}Formulas and computations}

\subsection{\label{appid}Tensor identities}

In the tensor identities provided below, $\mathbf{a}$ and $\mathbf{b}$\ denote
arbitrary vector functions, $\mathsf{M}$\ and $\mathsf{N}$\ are arbitrary
order $2$ tensor functions, and $\mathsf{Q}$\ is a real orthogonal order $2$
tensor that may be dependent on time. Furthermore, the superscripts, "$T$" and
"$SD$", and the symbol, "$\mathrm{tr}$", denote the transpose, symmetric
deviatoric part, and the trace of an order $2$\ tensor, respectively; and the
double-dot operator gives the inner tensor product computed between two order
$2$\ tensors as the scalar,%
\begin{equation}
\mathsf{M}\cdot\cdot\mathsf{N}=\sum_{i,j=1}^{3}M_{ij}N_{ij}.\label{t9}%
\end{equation}

The following is a list of identities used in Section \ref{obj} and Appendix
\ref{com}:%
\begin{equation}
\mathsf{M}^{SD}\cdot\cdot\mathsf{M}=\mathsf{M}^{SD}\cdot\cdot\mathsf{M}%
^{SD},\label{t2.6}%
\end{equation}%
\begin{equation}
\left(  \mathsf{Q}\cdot\mathbf{a}\right)  \cdot\left(  \mathsf{Q}%
\cdot\mathbf{b}\right)  =\mathbf{a}\cdot\mathbf{b}\label{p.1}%
\end{equation}%
\begin{equation}
\left(  \mathsf{Q}\cdot\mathsf{M}^{T}\right)  \cdot\cdot\left(  \mathsf{Q}%
\cdot\mathsf{M}^{T}\right)  =\mathsf{M}\cdot\cdot\mathsf{M},\label{p1.1}%
\end{equation}%
\begin{equation}
\left(  \mathsf{Q}\cdot\mathsf{M}\cdot\mathsf{Q}^{T}\right)  \cdot\cdot\left(
\mathsf{Q}\cdot\mathsf{M}\cdot\mathsf{Q}^{T}\right)  =\mathsf{M}\cdot
\cdot\mathsf{M},\label{p2}%
\end{equation}%
\begin{equation}
\mathrm{tr}\left(  \mathsf{Q}\cdot\mathsf{M}\cdot\mathsf{Q}^{T}\right)
=\mathrm{tr}\left(  \mathsf{M}\right)  ,\label{p3}%
\end{equation}%
\begin{equation}
\left(  \mathsf{Q}\cdot\mathsf{M}\cdot\mathsf{Q}^{T}\right)  \cdot\cdot\left(
\mathsf{Q}\cdot\mathsf{N}\right)  =\mathsf{M}\cdot\cdot\left(  \mathsf{N}%
\cdot\mathsf{Q}\right)  ,\label{p3.1}%
\end{equation}%
\begin{equation}
\mathsf{Q}\cdot\frac{d\mathsf{Q}^{T}}{dt}=-\frac{d\mathsf{Q}}{dt}%
\cdot\mathsf{Q}^{T},\label{p3.4}%
\end{equation}
and%
\begin{equation}
\mathrm{tr}\left(  \mathsf{Q}\cdot\frac{d\mathsf{Q}^{T}}{dt}\right)
=0.\label{p5}%
\end{equation}

\subsection{\label{com}Computations}

\begin{enumerate}
\item \label{com0}The rate of change of the kinetic energy density may be
computed as%
\begin{align*}
\frac{\partial\left(  \frac{1}{2}\rho\mathbf{v}\cdot\mathbf{v}\right)
}{\partial t}  & =\frac{1}{2}\frac{\partial}{\partial t}\left(  \frac{1}{\rho
}\rho\mathbf{v}\cdot\rho\mathbf{v}\right) \\
& =\frac{1}{2}\left(  \frac{\partial\left(  1/\rho\right)  }{\partial t}%
\rho^{2}\mathbf{v}\cdot\mathbf{v+}\frac{\partial\left(  \rho\mathbf{v}\right)
}{\partial t}\cdot\mathbf{v}+\mathbf{v}\cdot\frac{\partial\left(
\rho\mathbf{v}\right)  }{\partial t}\right)  ,
\end{align*}
which leads to%
\begin{equation}
\frac{\partial\left(  \frac{1}{2}\rho\mathbf{v}\cdot\mathbf{v}\right)
}{\partial t}=-\frac{1}{2}\mathbf{v}\cdot\mathbf{v}\frac{\partial\rho
}{\partial t}+\frac{\partial\left(  \rho\mathbf{v}\right)  }{\partial t}%
\cdot\mathbf{v}.\label{c.1}%
\end{equation}
By substituting the NSF mass and momentum conservation laws (\ref{a1}) and
(\ref{a2}) into the above, one finds%
\[
\frac{\partial\left(  \frac{1}{2}\rho\mathbf{v}\cdot\mathbf{v}\right)
}{\partial t}=\frac{1}{2}\mathbf{v}\cdot\mathbf{v}\nabla\cdot\left(
\rho\mathbf{v}\right)  -\left(  \nabla\cdot\left(  p\mathsf{I}-\mathsf{S}%
+\rho\mathbf{vv}\right)  \right)  \cdot\mathbf{v},
\]
or with the product rule,%
\begin{multline*}
\frac{\partial\left(  \frac{1}{2}\rho\mathbf{v}\cdot\mathbf{v}\right)
}{\partial t}=\nabla\cdot\left(  \left(  \frac{1}{2}\rho\mathbf{v}%
\cdot\mathbf{v}\right)  \mathbf{v}\right)  -\frac{1}{2}\rho\mathbf{v}%
\cdot\nabla\left(  \mathbf{v}\cdot\mathbf{v}\right) \\
-\nabla\cdot\left(  \left(  p\mathsf{I}-\mathsf{S}+\rho\mathbf{vv}\right)
\cdot\mathbf{v}\right)  +\left(  p\mathsf{I}-\mathsf{S}+\rho\mathbf{vv}%
\right)  \cdot\cdot\nabla\mathbf{v}.
\end{multline*}
One may employ additional algebra and calculus for tensors to rewrite some of
the terms appearing on the right-hand side as%
\begin{equation}
-\frac{1}{2}\rho\mathbf{v}\cdot\nabla\left(  \mathbf{v}\cdot\mathbf{v}\right)
=-\rho\mathbf{v}\cdot\left(  \nabla\mathbf{v}\right)  \cdot\mathbf{v}%
=-\rho\left(  \mathbf{vv}\right)  \cdot\cdot\nabla\mathbf{v},\label{a55}%
\end{equation}%
\begin{equation}
\rho\left(  \mathbf{vv}\right)  \cdot\mathbf{v}=\rho\left(  \mathbf{v}%
\cdot\mathbf{v}\right)  \mathbf{v},\label{a56}%
\end{equation}
and%
\begin{equation}
p\mathsf{I}\cdot\cdot\nabla\mathbf{v}=p\nabla\cdot\mathbf{v}.\label{a57}%
\end{equation}
Substitution of (\ref{a55})-(\ref{a57}) into our last expression for the rate
of change of kinetic energy density yields%
\[
\frac{\partial\left(  \frac{1}{2}\rho\mathbf{v}\cdot\mathbf{v}\right)
}{\partial t}=-\nabla\cdot\left(  -\mathsf{S}\cdot\mathbf{v}+\left(
p+\frac{1}{2}\rho\mathbf{v}\cdot\mathbf{v}\right)  \mathbf{v}\right)
+p\nabla\cdot\mathbf{v}-\mathsf{S}\cdot\cdot\nabla\mathbf{v}.
\]
Finally, by using constitutive relation (\ref{a6}) for the viscous stress
tensor with identity (\ref{t2.6}) to express the last term on the right-hand
side, we obtain the NSF kinetic energy balance law (\ref{a7})/(\ref{f2}).

\item \label{com1}Upon substitution of the dcE mass and momentum conservation
laws (\ref{f3}) and (\ref{f4}) into (\ref{c.1}), one finds%
\[
\frac{\partial\left(  \frac{1}{2}\rho\mathbf{v}\cdot\mathbf{v}\right)
}{\partial t}=\frac{1}{2}\left(  \mathbf{v}\cdot\mathbf{v}\right)  \nabla
\cdot\left(  -\nu\nabla\rho+\rho\mathbf{v}\right)  -\left(  \nabla\cdot\left(
p\mathsf{I}-\nu\nabla\left(  \rho\mathbf{v}\right)  +\rho\mathbf{vv}\right)
\right)  \cdot\mathbf{v}.
\]
Next, we carry out a procedure similar to the one used in the previous
computation for the NSF model. With the product rule, the above becomes%
\begin{multline*}
\frac{\partial\left(  \frac{1}{2}\rho\mathbf{v}\cdot\mathbf{v}\right)
}{\partial t}=\nabla\cdot\left(  \frac{1}{2}\rho\left(  \mathbf{v}%
\cdot\mathbf{v}\right)  \mathbf{v}\right)  -\frac{1}{2}\rho\mathbf{v}%
\cdot\nabla\left(  \mathbf{v}\cdot\mathbf{v}\right) \\
-\nabla\cdot\left(  \left(  p\mathsf{I}+\rho\mathbf{vv}\right)  \cdot
\mathbf{v}\right)  +\left(  p\mathsf{I}+\rho\mathbf{vv}\right)  \cdot
\cdot\nabla\mathbf{v}\\
+\left(  \nabla\cdot\left(  \nu\nabla\left(  \rho\mathbf{v}\right)  \right)
\right)  \cdot\mathbf{v}-\frac{1}{2}\left(  \mathbf{v}\cdot\mathbf{v}\right)
\nabla\cdot\left(  \nu\nabla\rho\right)  ,
\end{multline*}
or upon substitution of (\ref{a55})-(\ref{a57}),%
\[
\frac{\partial\left(  \frac{1}{2}\rho\mathbf{v}\cdot\mathbf{v}\right)
}{\partial t}=-\nabla\cdot\left(  \left(  p+\frac{1}{2}\rho\left(
\mathbf{v}\cdot\mathbf{v}\right)  \right)  \mathbf{v}\right)  +p\nabla
\cdot\mathbf{v}+\left(  \nabla\cdot\left(  \nu\nabla\left(  \rho
\mathbf{v}\right)  \right)  \right)  \cdot\mathbf{v}-\frac{1}{2}\left(
\mathbf{v}\cdot\mathbf{v}\right)  \nabla\cdot\left(  \nu\nabla\rho\right)  .
\]
Next, we may apply further tensor calculus and algebra to rewrite the last two
terms appearing on the right-hand side as%
\[
\left(  \nabla\cdot\left(  \nu\nabla\left(  \rho\mathbf{v}\right)  \right)
\right)  \cdot\mathbf{v}=\nabla\cdot\left(  \nu\nabla\left(  \rho
\mathbf{v}\cdot\mathbf{v}\right)  \right)  -\nabla\cdot\left(  \frac{1}{2}%
\rho\nu\nabla\left(  \mathbf{v}\cdot\mathbf{v}\right)  \right)  -\rho\nu
\nabla\mathbf{v}\cdot\cdot\nabla\mathbf{v}-\nu\nabla\rho\cdot\left(
\nabla\mathbf{v}\right)  \cdot\mathbf{v}%
\]
and%
\[
-\frac{1}{2}\left(  \mathbf{v}\cdot\mathbf{v}\right)  \nabla\cdot\left(
\nu\nabla\rho\right)  =\nabla\cdot\left(  -\frac{1}{2}\nu\left(
\mathbf{v}\cdot\mathbf{v}\right)  \nabla\rho\right)  +\nu\nabla\rho
\cdot\left(  \nabla\mathbf{v}\right)  \cdot\mathbf{v}.
\]
Substituting the above expressions into the previous equation for the rate of
change of kinetic energy density and combining terms with one more application
of the product rule yields the dcE kinetic energy balance law (\ref{f6.1}).

\item \label{com2}First, one observes that if the fixed coordinates,
$\mathbf{x}$, and the moving coordinates, $\widehat{\mathbf{x}}$, are related
via Euclidean transformation (\ref{f9}) and if the order $n$\ tensor, $A$, is
a local fluid property, then the chain rule implies%
\begin{equation}
\nabla_{\widehat{\mathbf{x}}}A=\mathsf{Q}\cdot\nabla A,\label{p1}%
\end{equation}
where $\nabla_{\widehat{\mathbf{x}}}$\ represents the gradient taken with
respect to the moving coordinates, $\widehat{\mathbf{x}}$. We can rewrite
Equation (\ref{f20}) as%
\begin{align*}
\widehat{\mathbf{v}}  & =\mathbf{v}\cdot\mathsf{Q}^{T}+\left(  \widehat
{\mathbf{x}}-\mathbf{o}\right)  \cdot\left(  \frac{d\mathsf{Q}}{dt}%
\cdot\mathsf{Q}^{T}\right)  ^{T}+\frac{d\mathbf{o}}{dt}\\
& =\mathbf{v}\cdot\mathsf{Q}^{T}+\left(  \widehat{\mathbf{x}}-\mathbf{o}%
\right)  \cdot\left(  \mathsf{Q}\cdot\frac{d\mathsf{Q}^{T}}{dt}\right)
+\frac{d\mathbf{o}}{dt}.
\end{align*}
Taking the gradient of the above and using (\ref{p1}), along with the product
rule and $\nabla_{\widehat{\mathbf{x}}}\widehat{\mathbf{x}}=\mathsf{I}$, then
yields%
\begin{equation}
\nabla_{\widehat{\mathbf{x}}}\widehat{\mathbf{v}}=\mathsf{Q}\cdot
\nabla\mathbf{v}\cdot\mathsf{Q}^{T}+\mathsf{Q}\cdot\frac{d\mathsf{Q}^{T}}%
{dt},\label{c2}%
\end{equation}
which is the velocity gradient in the moving reference frame, also denoted as
$\widehat{\nabla\mathbf{v}}$.

\item \label{com3}We take the trace of Equation (\ref{c2}) to find%
\begin{align*}
\nabla_{\widehat{\mathbf{x}}}\cdot\widehat{\mathbf{v}}  & =\mathrm{tr}\left(
\mathsf{Q}\cdot\nabla\mathbf{v}\cdot\mathsf{Q}^{T}\right)  +\mathrm{tr}\left(
\mathsf{Q}\cdot\frac{d\mathsf{Q}^{T}}{dt}\right) \\
& =\mathrm{tr}\left(  \nabla\mathbf{v}\right) \\
& =\nabla\cdot\mathbf{v},
\end{align*}
where properties (\ref{p3}) and (\ref{p5}) have been used in the second
line.\ The quantity, $\nabla_{\widehat{\mathbf{x}}}\cdot\widehat{\mathbf{v}}$,
is the divergence of the velocity in the moving reference frame, also denoted
as $\widehat{\nabla\cdot\mathbf{v}}$.

\item \label{com4}Let us use\ the superscript, "$S$", to denote the symmetric
part of an order $2$ tensor. From Equation (\ref{c2}), we compute the
symmetric part of the velocity gradient in the moving frame to be%
\begin{align*}
\nabla_{\widehat{\mathbf{x}}}\widehat{\mathbf{v}}^{S}  & =\frac{1}{2}\left(
\nabla_{\widehat{\mathbf{x}}}\widehat{\mathbf{v}}+\nabla_{\widehat{\mathbf{x}%
}}\widehat{\mathbf{v}}^{T}\right) \\
& =\left(
\begin{array}
[c]{c}%
\frac{1}{2}\left(  \mathsf{Q}\cdot\nabla\mathbf{v}\cdot\mathsf{Q}%
^{T}+\mathsf{Q}\cdot\nabla\mathbf{v}^{T}\cdot\mathsf{Q}^{T}\right)  +\\
\frac{1}{2}\left(  \mathsf{Q}\cdot\frac{d\mathsf{Q}^{T}}{dt}+\frac
{d\mathsf{Q}}{dt}\cdot\mathsf{Q}\right)
\end{array}
\right) \\
& =\mathsf{Q}\cdot\nabla\mathbf{v}^{S}\cdot\mathsf{Q}^{T}%
\end{align*}
by property (\ref{p3.4}), and from this one observes the symmetric part of the
velocity gradient to be objective. Next, the deviatoric part of the above is
computed as%
\begin{align*}
\nabla_{\widehat{\mathbf{x}}}\widehat{\mathbf{v}}^{SD}  & =\nabla
_{\widehat{\mathbf{x}}}\widehat{\mathbf{v}}^{S}-\frac{1}{3}\mathrm{tr}\left(
\nabla_{\widehat{\mathbf{x}}}\widehat{\mathbf{v}}^{S}\right)  \mathsf{I}\\
& =\mathsf{Q}\cdot\nabla\mathbf{v}^{S}\cdot\mathsf{Q}^{T}-\frac{1}%
{3}\mathrm{tr}\left(  \mathsf{Q}\cdot\nabla\mathbf{v}^{S}\cdot\mathsf{Q}%
^{T}\right)  \mathsf{I}\\
& =\mathsf{Q}\cdot\nabla\mathbf{v}^{S}\cdot\mathsf{Q}^{T}-\frac{1}%
{3}\mathrm{tr}\left(  \nabla\mathbf{v}^{S}\right)  \mathsf{I}\\
& =\mathsf{Q}\cdot\left(  \nabla\mathbf{v}^{S}-\frac{1}{3}\mathrm{tr}\left(
\nabla\mathbf{v}^{S}\right)  \mathsf{I}\right)  \cdot\mathsf{Q}^{T}\\
& =\mathsf{Q}\cdot\nabla\mathbf{v}^{SD}\cdot\mathsf{Q}^{T},
\end{align*}
where we have used property (\ref{p3}) in the third line. The quantity,
$\nabla_{\widehat{\mathbf{x}}}\widehat{\mathbf{v}}^{SD}$, is the symmetric
deviatoric part of the velocity gradient in the moving reference frame, also
denoted as $\widehat{\nabla\mathbf{v}^{SD}}$.
\end{enumerate}

\end{document}